\documentclass{article}
\usepackage{spconf,amsmath,graphicx}

\newcommand{\bi}[1]{\ensuremath{\boldsymbol{#1}}}   

\usepackage{amssymb}
\usepackage{enumerate}
\usepackage{graphicx}
\usepackage{multirow}
\usepackage{arydshln}
\usepackage{color}
\usepackage{float}
\usepackage{bm}
\usepackage{comment}
\usepackage{cite}
\usepackage{url}
\usepackage{amsmath}
\usepackage{algorithmic}
\usepackage{array}
\usepackage{colortbl}

\newlength\savedwidth
\newcommand{\wcline}[1]{\noalign{\global\savedwidth\arrayrulewidth\global\arrayrulewidth 1.0pt} \cline{#1}
\noalign{\global\arrayrulewidth\savedwidth}}

\title{Sound Event Detection by Multitask Learning of\\Sound Events and Scenes with Soft Scene Labels}
\name{Keisuke Imoto\hspace{1pt}$^{1 \ast}$, Noriyuki Tonami\hspace{1pt}$^{1 \ast}$, Yuma Koizumi\hspace{1pt}$^{2}$, Masahiro Yasuda\hspace{1pt}$^{2}$,
}
\secondlinename{Ryosuke Yamanishi\hspace{1pt}$^{1}$, and Yoichi Yamashita\hspace{1pt}$^{1}$
}
\address{$^{1}$ Ritsumeikan University, Shiga, Japan, $^{2}$ NTT Media Intelligence Laboratories, Tokyo, Japan}
\begin{document}
%
\maketitle
%
\begin{abstract}
Sound event detection (SED) and acoustic scene classification (ASC) are major tasks in environmental sound analysis. Considering that sound events and scenes are closely related to each other, some works have addressed joint analyses of sound events and acoustic scenes based on multitask learning (MTL), in which the knowledge of sound events and scenes can help in estimating them mutually. The conventional MTL-based methods utilize one-hot scene labels to train the relationship between sound events and scenes; thus, the conventional methods cannot model the extent to which sound events and scenes are related. However, in the real environment, common sound events may occur in some acoustic scenes; on the other hand, some sound events occur only in a limited acoustic scene. In this paper, we thus propose a new method for SED based on MTL of SED and ASC using the soft labels of acoustic scenes, which enable us to model the extent to which sound events and scenes are related. Experiments conducted using TUT Sound Events 2016/2017 and TUT Acoustic Scenes 2016 datasets show that the proposed method improves the SED performance by 3.80\% in F-score compared with conventional MTL-based SED.
%
\end{abstract}
%
\vspace{-3pt}
%
%
\vspace{-6pt}
\section{Introduction}
\label{sec:intro}
\vspace{-7pt}
\renewcommand{\thefootnote}{\fnsymbol{footnote}}
\footnotetext[1]{These authors contributed equally to this work.}
Environmental sound analysis has great potential for developing many applications such as monitoring systems \cite{Peng_ICME2009_01}, abnormal sound detection systems \cite{Chan_EUSIPCO2010_01,Koizumi_TASLP2019_01}, automatic surveillance \cite{Radhakrishnan_WASPAA2005_01,Harma_ICME2005_01,Ntalampiras_ICASSP2009_01}, and media retrieval \cite{Jin_INTERSPEECH2012_01}.
For environmental sound analysis, sound event detection (SED) and acoustic scene classification (ASC) have mainly been studied.
SED involves detecting sound event labels and their onset/offset in an audio recording, where a sound event indicates a type of sound such as ``mouse clicking,'' ``people talking,'' or ``bird singing.''
ASC involves predicting acoustic scene labels in an audio recording, where an acoustic scene indicates a recording situation, place, or human activity such as ``office,'' ``train,'' or ``cooking.''

\ \\[-12pt]
\indent
In particular, many SED methods based on the Gaussian mixture model (GMM) \cite{Xiaodan_ICASSP2009_01} and hidden Markov model (HMM) \cite{Mesaros_EUSIPCO2010_01} have been proposed.
However, these approaches cannot detect multiple overlapping sound events; thus, polyphonic SED systems have also been developed.
One approach to polyphonic SED is the use of non-negative matrix factorization (NMF) \cite{Komatsu_DCASE2016_01}.
More recently, polyphonic SED systems based on neural networks have also been developed \cite{Hershey_ICASSP2017_01,Cakir_TASLP2017_01,Hayashi_TASLP2017_01}.
For example, Hershey \textit{et al.} have proposed an event detection method based on a convolutional neural network (CNN) \cite{Hershey_ICASSP2017_01}.
\c{C}ak\i r \textit{et al.} \cite{Cakir_TASLP2017_01} and Hayashi \textit{et al.} \cite{Hayashi_TASLP2017_01} have proposed methods utilizing a recurrent neural network (RNN) or a convolutional recurrent neural network (CRNN), which can capture temporal information of sound events.

\begin{table*}[t]
\small
\caption{Sound events occurring in each acoustic scene in TUT Acoustic Scenes 2016 [23] and 2017 [24]}
\label{tbl:event_occurrence}
\centering
\begin{tabular}{c}
\hspace{-7pt}
\includegraphics[width=2.07\columnwidth]{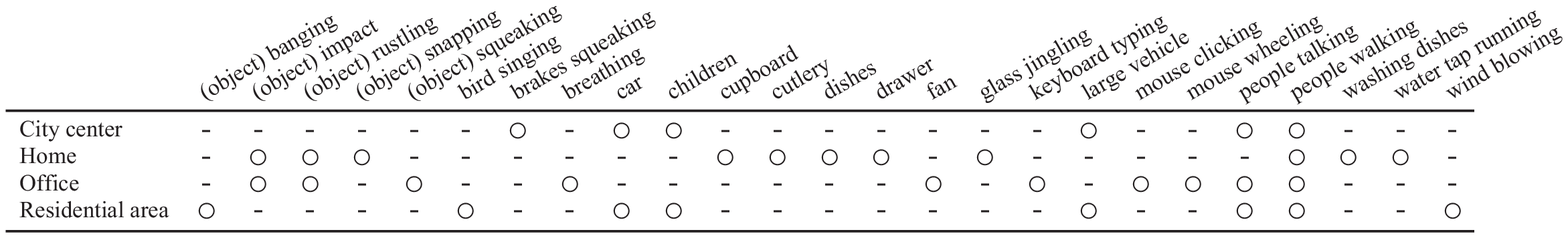}
\end{tabular}
\vspace{-9pt}
\end{table*}
%

Sound events and scenes are related to each other; for instance, in an acoustic scene ``office,'' the sound events ``mouse clicking,'' and ``keyboard typing'' tend to occur, whereas the sound events ``large vehicle'' and ``bird singing'' are not likely to occur.
Thus, when analyzing the sound events ``mouse clicking'' and ``keyboard typing,'' information on the acoustic scene ``office'' will help in detecting these sound events, and vice versa.
On the basis of this idea, Mesaros \textit{et al.} \cite{Mesaros_EUSIPCO2011_01} and Heittola \textit{et al.} \cite{Heittola_JASM2013_01} have proposed the SED method utilizing information on acoustic scenes in an unsupervised manner, and Imoto and coworkers \cite{Imoto_IEICE2016_01,Imoto_TASLP2019_01} have proposed ASC taking information on sound events into account, which is based on Bayesian generative models.
Bear \textit{et al.} \cite{Bear_INTERSPEECH2019_01} and Tonami \textit{et al.} \cite{Tonami_WASPAA2019_01} have proposed joint analyzing methods of sound events and scenes based on multitask learning (MTL) of SED and ASC, in which one-hot scene labels is used to train the models.

In the real environment, common sound events may occur in some acoustic scenes, for example, in the acoustic scenes ``residential area'' and ``city center,'' the common sound events ``car'' may occur.
On the other hand, some sound events occur only in a limited acoustic scene, for example, the sound event ``dishes'' occurs in the acoustic scene ``home'' in most cases.
This indicates that there is the extent to which sound events and scenes are related.
However, the conventional MTL-based methods cannot consider the extent of the relevance because the methods utilize one-hot scene labels to train the models.
To overcome this problem, we propose a new MTL-based SED method with soft scene labels, which is obtained using the teacher--student learning framework.
%
%
%
%
%
%
\vspace{-4pt}
\section{Conventional Methods}
\vspace{-9pt}
\subsection{Conventional Methods for Event Detection and Scene Classification}
\label{ssec:conventional}
\vspace{-3pt}
In this section, we introduce the conventional SED and ASC methods.
Recently, many neural network-based methods, such as convolutional neural network (CNN)-based methods \cite{Hershey_ICASSP2017_01,Sakashita_DCASE2018_01} and a recurrent neural network (RNN)-based method \cite{Hayashi_TASLP2017_01}, have been proposed.
%
For an example in CNN-based SED and ASC, the time-frequency representation of the observed signal ${\bf V} \in \mathbb{R}^{D \times N}$, such as log-mel-band energy, is fed to a convolutional layer, where $D$ and $N$ are the number of frequency bins and the number of time frames of the input feature, respectively.
In the convolution layer, the input feature map is convoluted with two-dimensional filters, then max pooling is conducted to reduce the dimension of the feature map.
The CNN architecture allows robust feature extraction against time and frequency shifts, which frequently occur in environmental sounds.
The output of the convolution layer is then input to the fully connected layer, which is followed by the sigmoid function for SED or softmax function for ASC.

SED involves the estimation of sound event labels and their onset/offset times, where acoustic events may overlap in the time axis.
Thus, the network for SED is optimized under the following sigmoid cross-entropy objective function $E_{1}({\bi \Theta}_{1})$:
\vspace{-6pt}
\begin{align}
E_{1} ({\bi \Theta}_{1}) &= - \! \sum^{N}_{n=1} \! {\big \{} {\bf z}_{n} \log {\big (} s({\bf y}_{n}) {\big )} \! + \! (1-{\bf z}_{n}) \log {\big (}1-s({\bf y}_{n}){\big )} \! {\big \}} \nonumber\\[-2pt]
&= - \sum^{M}_{m=1} \sum^{N}_{n=1} \! {\Big \{} z_{m,n} \log {\big (} s(y_{m,n}) {\big )} \nonumber\\[-4pt]
&\hspace{25pt} + (1 - z_{m,n}) \log {\big (} 1 - s(y_{m,n}) {\big )} {\Big \}},
\label{eq:event_loss}
\end{align}
\vspace{-11pt}

\noindent where $s$, $M$, $y_{m,n}$, and $z_{m,n}$ are, respectively, the sigmoid function, the number of the acoustic event category, the output of the fully connected layer in time frame $n$, and the target label in time frame $n$, which is 1 if acoustic event $m$ is active in time frame $n$, and 0 otherwise.

On the other hand, ASC involves the estimation of the acoustic scene label with which a sound clip is most associated.
The network for ASC is optimized under the following softmax cross-entropy objective function $E_{2}({\bi \Theta}_{2})$:
\vspace{-5pt}
\begin{align}
E_{2}({\bi \Theta}_{2}) = - \sum^{C}_{c=1} {\Big \{} z_{c} \log {\big (} \sigma (y_{c}) {\big )} {\Big \}},
\label{eq:scene_hard_loss}
\end{align}
\vspace{-10pt}

\noindent where $\sigma$, $C$, $y_c$, and $z_c$ are the softmax function, the number of acoustic scene categories, the output of the fully connected layer, and the hard scene label, respectively.
%
%
%
%
%
\begin{figure*}[t!]
\centering
\begin{tabular}{c}
\hspace{-8pt}
\begin{minipage}{0.375\hsize}
\centering
\includegraphics[width=0.96\columnwidth]{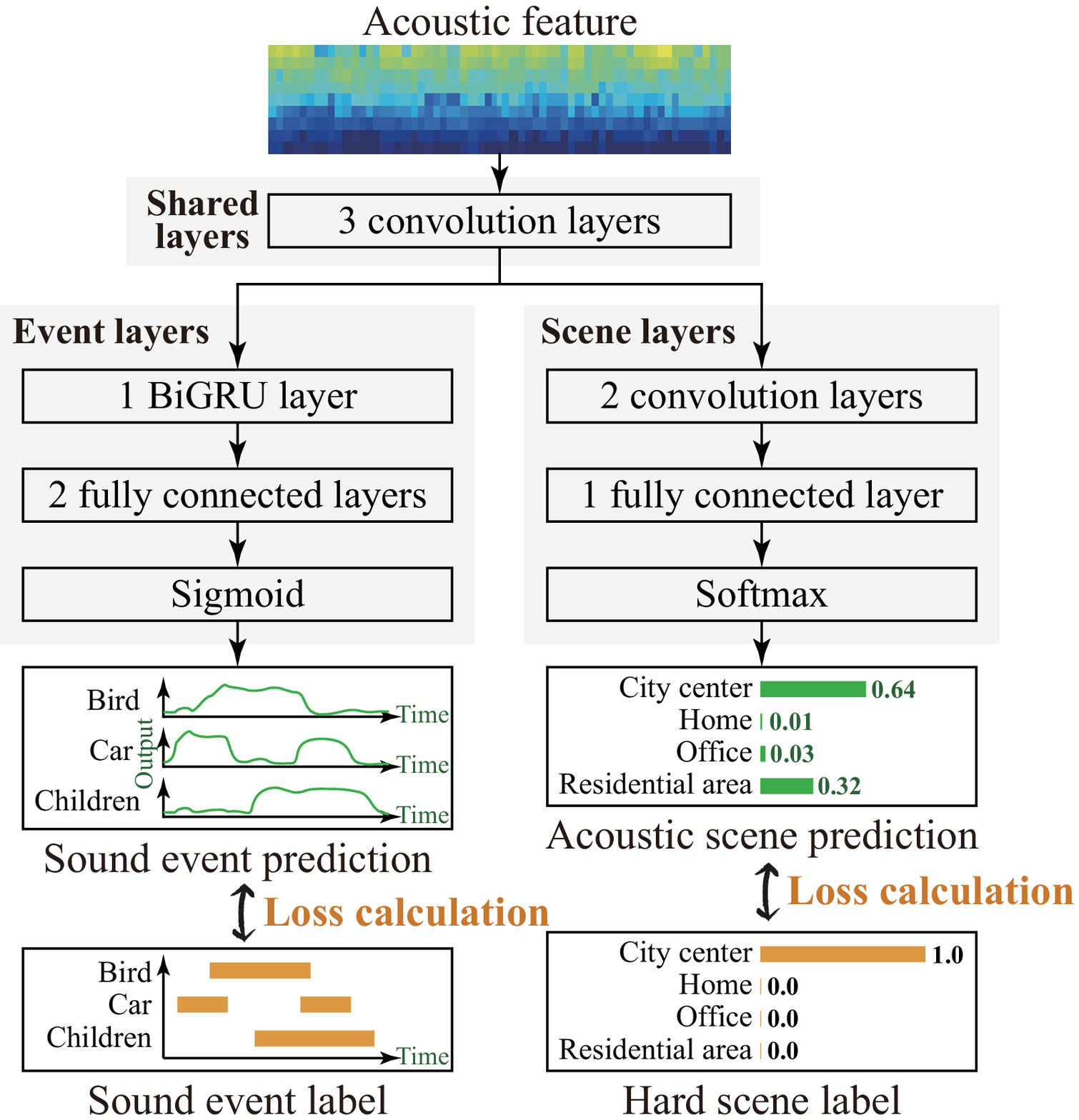}
\vspace{-8pt}
\caption{Network structure of conventional MTL-based method \cite{Tonami_WASPAA2019_01}}
\label{fig:convnet}
\end{minipage}
\hspace{4pt}
\begin{minipage}{0.615\hsize}
\centering
\includegraphics[width=0.96\columnwidth]{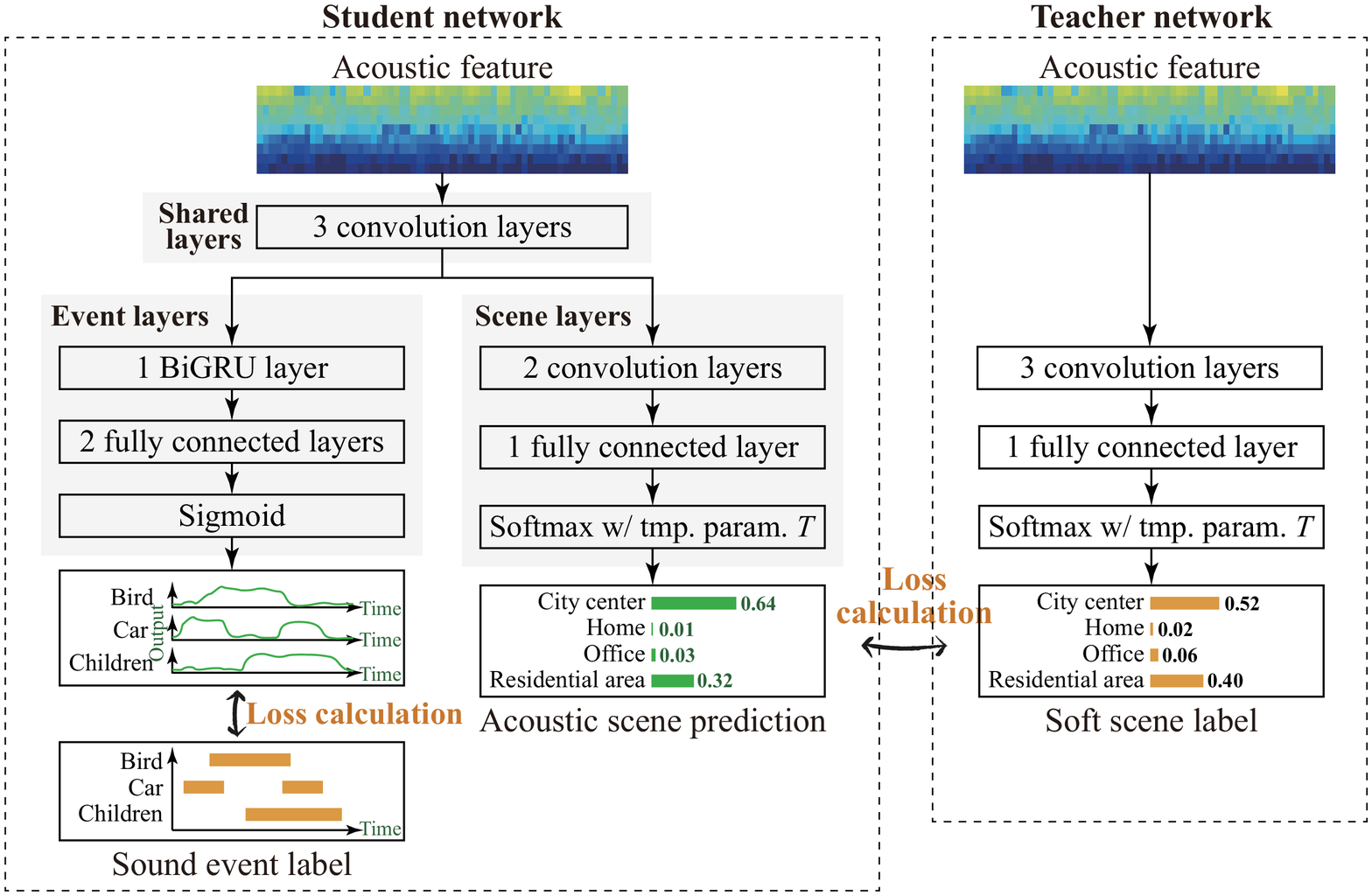}
\vspace{-9pt}
\caption{Proposed MTL-based SED with soft scene label}
\label{fig:propnet}
\end{minipage}
\end{tabular}
\vspace{-8pt}
\end{figure*}
%
%
%
%
\vspace{-8pt}
\subsection{Joint Analysis of Sound Events and Scenes Based on Multitask Learning}
\label{ssec:MTL}
\vspace{-2pt}
Most works address SED and ASC separately.
However, some sound events and scenes are closely related, and the knowledge of sound events and scenes can help in estimating them mutually.
Considering this idea, joint analysis of sound events and acoustic scenes based on MTL of SED and ASC has been proposed \cite{Bear_INTERSPEECH2019_01,Tonami_WASPAA2019_01}.
As shown in Fig.~\ref{fig:convnet}, these methods share parts of the networks holding information on sound events and scenes in common.

To optimize a multitask-based network, the following objective function is used in \cite {Tonami_WASPAA2019_01}:
%
\begin{align}
E({\bi \Theta}) = E_{1}({\bi \Theta}_{1}) + \alpha E_{2}({\bi \Theta}_{2}),
\label{eq:convloss}
\end{align}
\vspace{-11pt}

\noindent where $\alpha$ indicates the weight of the scene loss.
In particular, the previous work \cite{Tonami_WASPAA2019_01} showed that an MTL-based method achieves a better performance in detecting sound events than CRNN-based SED \cite{Cakir_TASLP2017_01}.
%
%
\vspace{-5pt}
\section{SED Based on Multitask Learning with Scene Soft Labels}
\label{sec:proposed}
\vspace{-7pt}
\subsection{Motivation}
\label{ssec:motivation}
\vspace{-2pt}
%
%
%
As shown in Table~\ref{tbl:event_occurrence}, common sound events may occur in some acoustic scenes, for example, in the acoustic scenes ``residential area'' and ``city center,'' the common sound events ``car'' and/or ``people walking'' may occur.
On the other hand, some sound events occur only in a limited acoustic scene, for example, the sound event ``dishes'' occurs only in the acoustic scene ``home.''
This means that there is the extent to which sound events and scenes are related.
However, the conventional MTL-based method utilizes one-hot (hard) scene labels to train the relationship between sound events and scenes; therefore, the conventional method cannot model the extent to which sound events and scenes are related.
To address this limitation, we propose SED by MTL of sound events and scenes with soft scene labels obtained by the teacher--student learning framework.
%
%
%
\vspace{-8pt}
\subsection{Proposed Method}
\label{ssec:prop}
\vspace{-5pt}
An overview of the proposed method is shown in Fig.~\ref{fig:propnet}.
In the proposed method, we adopt the teacher--student learning framework \cite{Hinton_NIPSDLRPWorkshop2015_01,Heo_INTERSPEECH2019_01} to the MTL-based model.
In the proposed scheme, the teacher network for scene classification is first trained using hard scene labels as in the conventional ASC method.
Then, the output of the trained teacher network is used as the soft scene label for training the student network.

To optimize parameters related to acoustic scenes in the student network, the following objective function is used:
%
\begin{align}
{\bf q}_{c} &= \frac{\exp (w_{c}/T)}{\sum_{i=1}^{C} \exp (w_{i}/T)},\\
{\bf p}_{c} &= \frac{\exp (v_{c}/T)}{\sum_{i=1}^{C} \exp (v_{i}/T)},\\
E_{3}({\bi \Theta}_{3}) &= -\sum_{c=1}^{C} ({\bf p}_{c} \log {\bf q}_{c}),
\label{eq:qn}
\end{align}
\vspace{-10pt}

\noindent where $w$ and $v$ are the outputs of fully connected layers of student and teacher models, respectively.
$T$ is the temperature parameter, which determines the extent of soft-label utilization \cite{Hinton_NIPSDLRPWorkshop2015_01}.
In the proposed method, the hard scene loss $E_{2}({\bi \Theta}_{2})$ in Eq.~\ref{eq:convloss} is replaced with $E_{3}({\bi \Theta}_{3})$.
Thus, the objective function of the proposed method is finally represented as
%
%
\begin{align}
E({\bi \Theta}) = E_{1}({\bi \Theta}_{1}) + \beta E_{3}({\bi \Theta}_{3}),
\label{eq:proploss}
\end{align}
\vspace{-15pt}

\noindent where $\beta$ is the weight of the soft scene loss.
%
%
%
%
%
\begin{table}[t]
\vspace{-12pt}
\small
\caption{Experimental conditions}
\label{tbl:parameter}
\begin{center}
\begin{tabular}{ll}
\wcline{1-2}
\!\!&\!\!\\[-10pt]
\!Structure of teacher network \!\!&\!\! 3 CNN \& 1 fully conn.\!\\[-1pt]
\!\# channels of CNN layers (shared) \!\!&\!\! 128, 128, 128\!\\[-1pt]
\!Filter size (shared) \!\!&\!\! 3$\times$3\!\\[-1pt]
\!Pooling size (shared) \!\!&\!\! 8\!$\times$\!8,\! 4\!$\times$\!4,\! 2\!$\times$\!2 (max pooling)\!\\[-1pt]
\cline{1-2}
\!&\\[-10pt]
\!Network structure of shared layers \!\!&\!\! 3 CNN\!\\[-1pt]
\!\# channels of CNN layers (shared) \!\!&\!\! 128, 128, 128\!\\[-1pt]
\!Filter size (shared) \!\!&\!\! 3$\times$3\!\\[-1pt]
\!Pooling size (shared) \!\!&\!\! 1\!$\times$\!8,\! 1\!$\times$\!4,\! 1\!$\times$\!2 (max pooling)\!\\[-1pt]
\cline{1-2}
\!&\\[-10pt]
\!Network structure of scene layers \!\!&\!\! 2 CNN\!\\[-1pt]
\!\# channels of CNN layers (scene) \!\!&\!\! 64, 16\!\\[-1pt]
\!Filter size (scene) \!\!&\!\! 3$\times$3\!\\[-1pt]
\!Pooling size (scene) \!\!&\!\! 10\!$\times$\!1,\! 5\!$\times$\!1 (max pooling)\!\\[-1pt]
\cline{1-2}
\!&\\[-10pt]
\!Network structure of event layers \!\!&\!\! 1 BiGRU \& 1 fully conn.\!\\[-1pt]
\!\# units in GRU layer (event) \!\!&\!\! 32\!\\[-1pt]
\!\# units in fully conn. layer (event) \!\!&\!\! 32\!\\[-1pt]
\wcline{1-2}
\end{tabular}
\vspace{-14pt}
\end{center}
\end{table}
\begin{table}[t]
\vspace{-8pt}
\caption{Performance of sound event detection}
\vspace{2pt}
\small
\centering
\begin{tabular}{lcc}
\wcline{1-3}
&&\\[-10pt]
\multicolumn{1}{c}{Method} & F-score & ER\\[-1pt]
\wcline{1-3}
&&\\[-9pt]
CNN-BiGRU & 42.17\% & 0.756\\[-1pt]
MTL ($\alpha$=0.0001) & 46.02\% & 0.724 \\[-1pt]
MTL ($\alpha$=1.0) & 29.26\% & 0.837 \\[-1pt]
MTL w/ soft labels ($\beta$=1.0, T=1.0) & \textbf{49.82\%} & \textbf{0.691} \\[-1pt]
\wcline{1-3}
\end{tabular}
\vspace{-3pt}
\label{tbl:performance01}
\end{table}
\begin{figure}[t]
\vspace{-6pt}
\centering
\includegraphics[width=0.87\columnwidth]{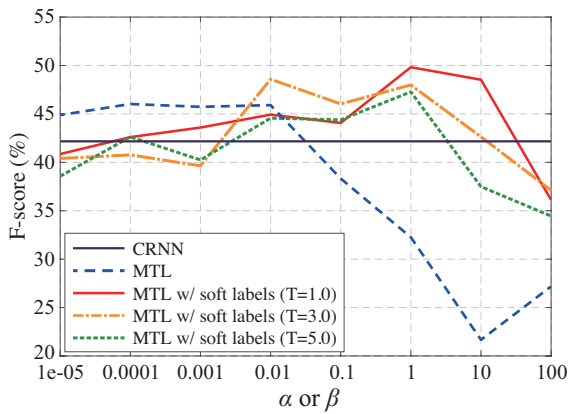}
\vspace{-11pt}
\caption{Sound event detection performance with various weights $\alpha$ or $\beta$}
\label{fig:performance}
\vspace{-9pt}
\end{figure}
\begin{table*}[t]
\vspace{-3pt}
\small
\caption{Sound event detection performance for each event}
\label{tbl:each_event}
\centering
\scalebox{0.779}[0.779]{
\begin{tabular}{llcccccccccccc}
\wcline{1-14}
&&&&&&&&&&&&&\\[-10pt]
\multicolumn{2}{c}{\multirow{2}{*}{Event}} & (object) & (object) & (object) & (object) & (object) & bird & brakes & \multirow{2}{*}{breathing} & \multirow{2}{*}{car} & \multirow{2}{*}{children} & \multirow{2}{*}{cupboard} & \multirow{2}{*}{cutlery}\\[-1pt]
&& banging & impact & rustling & snapping &  squeaking & singing & squeaking &&&&&\\[0pt]
\wcline{1-14}\\[-10pt]
\multirow{2.2}{*}{CRNN}& \cellcolor[rgb]{0.95,0.95,0.95} F-score & \cellcolor[rgb]{0.95,0.95,0.95} 0.00\% & \cellcolor[rgb]{0.95,0.95,0.95} 1.20\% & \cellcolor[rgb]{0.95,0.95,0.95} \textbf{0.16\%} & \cellcolor[rgb]{0.95,0.95,0.95} 0.08\% & \cellcolor[rgb]{0.95,0.95,0.95} 0.00\% & \cellcolor[rgb]{0.95,0.95,0.95} 25.14\% & \cellcolor[rgb]{0.95,0.95,0.95} 3.00\% & \cellcolor[rgb]{0.95,0.95,0.95} 0.00\% & \cellcolor[rgb]{0.95,0.95,0.95} 58.72\% & \cellcolor[rgb]{0.95,0.95,0.95} 0.00\% & \cellcolor[rgb]{0.95,0.95,0.95} \textbf{2.31\%} & \cellcolor[rgb]{0.95,0.95,0.95} 0.47\%\\[0pt]
\cline{2-14}\\[-9.8pt]
& Error rate & 0.0011 & 0.0263 & 0.0247 & 0.0018 & 0.0015 & 0.1022 & 0.0105 & 0.0019 & 0.1469 & 0.0255 & 0.0017 & 0.0040\\[0pt]
\cline{1-14}\\[-10.5pt]
\multirow{2.2}{*}{MTL ($\alpha = 0.0001$)} & \cellcolor[rgb]{0.95,0.95,0.95} F-score & \cellcolor[rgb]{0.95,0.95,0.95} 0.00\% & \cellcolor[rgb]{0.95,0.95,0.95} \textbf{2.26\%} & \cellcolor[rgb]{0.95,0.95,0.95} 0.14\% & \cellcolor[rgb]{0.95,0.95,0.95} 0.08\% & \cellcolor[rgb]{0.95,0.95,0.95} 0.00\% & \cellcolor[rgb]{0.95,0.95,0.95} \textbf{44.91\%} & \cellcolor[rgb]{0.95,0.95,0.95} \textbf{5.95\%} & \cellcolor[rgb]{0.95,0.95,0.95} 0.00\% & \cellcolor[rgb]{0.95,0.95,0.95} \textbf{59.48\%} & \cellcolor[rgb]{0.95,0.95,0.95} 0.00\% & \cellcolor[rgb]{0.95,0.95,0.95} 0.08\% & \cellcolor[rgb]{0.95,0.95,0.95} 0.17\% \\[0pt]
\cline{2-14}\\[-10pt]
& Error rate & 0.0011 & \textbf{0.0262} & 0.0247 & 0.0018 & 0.0015 & 0.0903 & 0.0104 & 0.0019 & 0.1493 & 0.0255 & 0.0017 & 0.0040\\[0pt]
\cline{1-14}\\[-10.5pt]
\multirow{2.2}{*}{MTL ($\alpha = 1.0$)} & \cellcolor[rgb]{0.95,0.95,0.95} F-score & \cellcolor[rgb]{0.95,0.95,0.95} 0.00\% & \cellcolor[rgb]{0.95,0.95,0.95} 0.03\% & \cellcolor[rgb]{0.95,0.95,0.95} 0.01\% & \cellcolor[rgb]{0.95,0.95,0.95} 0.00\% & \cellcolor[rgb]{0.95,0.95,0.95} 0.00\% & \cellcolor[rgb]{0.95,0.95,0.95} 0.00\% & \cellcolor[rgb]{0.95,0.95,0.95} 0.59\% & \cellcolor[rgb]{0.95,0.95,0.95} 0.00\% & \cellcolor[rgb]{0.95,0.95,0.95} 43.46\% & \cellcolor[rgb]{0.95,0.95,0.95} 0.00\% & \cellcolor[rgb]{0.95,0.95,0.95} 0.00\% & \cellcolor[rgb]{0.95,0.95,0.95} \textbf{2.58\%} \\\cline{2-14}\\[-10pt]
& Error rate & 0.0011 & 0.0263 & 0.0247 & 0.0018 & 0.0015 & 0.1134 & 0.0106 & 0.0019 & 0.1776 & 0.0255 & 0.0017 & \textbf{0.0039}\\[0pt]
\cline{1-14}\\[-10.5pt]
\multirow{1.1}{*}{MTL w/ soft labels}& \cellcolor[rgb]{0.95,0.95,0.95} F-score & \cellcolor[rgb]{0.95,0.95,0.95} 0.00\% & \cellcolor[rgb]{0.95,0.95,0.95} 0.53\% & \cellcolor[rgb]{0.95,0.95,0.95} 0.02\% & \cellcolor[rgb]{0.95,0.95,0.95} 0.00\% & \cellcolor[rgb]{0.95,0.95,0.95} 0.00\% & \cellcolor[rgb]{0.95,0.95,0.95} 43.93\% & \cellcolor[rgb]{0.95,0.95,0.95} 5.13\% & \cellcolor[rgb]{0.95,0.95,0.95} 0.00\% & \cellcolor[rgb]{0.95,0.95,0.95} 58.87\% & \cellcolor[rgb]{0.95,0.95,0.95} 0.00\% & \cellcolor[rgb]{0.95,0.95,0.95} 1.12\% & \cellcolor[rgb]{0.95,0.95,0.95} 0.03\% \\\cline{2-14}
\multirow{0.9}{*}{($\beta = 1.0, T = 1.0$)}&&&&&&&&&&&&&\\[-10pt]
& Error rate & 0.0011 & 0.0263 & 0.0247 & 0.0018 & 0.0015 & \textbf{0.0867} & \textbf{0.0103} & 0.0019 & \textbf{0.1467} & 0.0255 & 0.0017 & 0.0040\\[-1pt]
\wcline{1-14}
\vspace{-5pt}
\end{tabular}
}
%
%
\small
\centering
\scalebox{0.7665}[0.7665]{
\begin{tabular}{llccccccccccccc}
\wcline{1-15}
&&&&&&&&&&&&&\\[-10pt]
\multicolumn{2}{c}{\multirow{2}{*}{Event}} &\! \multirow{2}{*}{dishes} \!\!&\!\! \multirow{2}{*}{drawer} \!\!&\!\! \multirow{2}{*}{fan} \!\!&\!\! glass \!\!&\!\! keyboard \!\!&\!\! large \!\!&\!\! mouse \!\!&\!\! mouse \!\!&\!\! people \!\!&\!\! people \!\!&\!\! washing \!\!&\!\! water tap \!\!&\!\! wind\!\!\\[-1pt]
&&\!\!  \!\!&\!\!  \!\!&\!\!  \!\!&\!\! jingling \!\!&\!\! typing \!\!&\!\! vehicle \!\!&\!\! clicking \!\!&\!\! wheeling \!\!&\!\! talking \!\!&\!\! walking \!\!&\!\! dishes \!\!&\!\! running \!\!&\!\! blowing\!\\[0pt]
\wcline{1-15}\\[-10pt]
\multirow{2.2}{*}{CRNN} & \cellcolor[rgb]{0.95,0.95,0.95} F-score &\! \cellcolor[rgb]{0.95,0.95,0.95} 0.19\% \!\!&\!\! \cellcolor[rgb]{0.95,0.95,0.95} 0.00\% \!\!&\!\! \cellcolor[rgb]{0.95,0.95,0.95} 62.18\% \!\!&\!\! \cellcolor[rgb]{0.95,0.95,0.95} 0.00\% \!\!&\!\! \cellcolor[rgb]{0.95,0.95,0.95} 0.61\% \!\!&\!\! \cellcolor[rgb]{0.95,0.95,0.95} 41.69\% \!\!&\!\! \cellcolor[rgb]{0.95,0.95,0.95} 0.00\% \!\!&\!\! \cellcolor[rgb]{0.95,0.95,0.95} 0.00\% \!\!&\!\! \cellcolor[rgb]{0.95,0.95,0.95} 0.05\% \!\!&\!\! \cellcolor[rgb]{0.95,0.95,0.95} 46.35\% \!\!&\!\! \cellcolor[rgb]{0.95,0.95,0.95} 10.73\% \!\!&\!\! \cellcolor[rgb]{0.95,0.95,0.95} \textbf{48.27\%} \!\!&\!\! \cellcolor[rgb]{0.95,0.95,0.95} 0.00\% \!\\\cline{2-15}\\[-10pt]
& Error rate &\! 0.0114 \!\!&\!\! 0.0018 \!\!&\!\! 0.1665 \!\!&\!\! 0.0020 \!\!&\!\! 0.0248 \!\!&\!\! \textbf{0.0593} \!\!&\!\! 0.0069 \!\!&\!\! 0.0024 \!\!&\!\! 0.0988 \!\!&\!\! 0.1042 \!\!&\!\! 0.0244 \!\!&\!\! \textbf{0.0155} \!\!&\!\! 0.0134 \!\\
\cline{1-15}\\[-10.3pt]
\multirow{2.2}{*}{MTL ($\alpha = 0.0001$)} & \cellcolor[rgb]{0.95,0.95,0.95} F-score &\! \cellcolor[rgb]{0.95,0.95,0.95} 0.80\% \!\!&\!\! \cellcolor[rgb]{0.95,0.95,0.95} 0.00\% \!\!&\!\! \cellcolor[rgb]{0.95,0.95,0.95} 60.55\% \!\!&\!\! \cellcolor[rgb]{0.95,0.95,0.95} 0.00\% \!\!&\!\! \cellcolor[rgb]{0.95,0.95,0.95} 3.91\% \!\!&\!\! \cellcolor[rgb]{0.95,0.95,0.95} 39.09\% \!\!&\!\! \cellcolor[rgb]{0.95,0.95,0.95} 0.00\% \!\!&\!\! \cellcolor[rgb]{0.95,0.95,0.95} 0.00\% \!\!&\!\! \cellcolor[rgb]{0.95,0.95,0.95} \textbf{1.13\%} \!\!&\!\! \cellcolor[rgb]{0.95,0.95,0.95} 48.78\% \!\!&\!\! \cellcolor[rgb]{0.95,0.95,0.95} \textbf{13.51\%} \!\!&\!\! \cellcolor[rgb]{0.95,0.95,0.95} 44.11\% \!\!&\!\! \cellcolor[rgb]{0.95,0.95,0.95} 0.00\% \!\\
\cline{2-15}\\[-10pt]
& Error rate &\! 0.0114 \!\!&\!\! 0.0018 \!\!&\!\! 0.1570 \!\!&\!\! 0.0020 \!\!&\!\! 0.0247 \!\!&\!\! 0.0623 \!\!&\!\! 0.0069 \!\!&\!\! 0.0024 \!\!&\!\! 0.0987 \!\!&\!\! \textbf{0.1019} \!\!&\!\! \textbf{0.0241} \!\!&\!\! 0.0161 \!\!&\!\! 0.0134 \!\\
\cline{1-15}\\[-10.3pt]
\multirow{2.2}{*}{MTL ($\alpha = 1.0$)} & \cellcolor[rgb]{0.95,0.95,0.95} F-score &\! \cellcolor[rgb]{0.95,0.95,0.95} 0.00\% \!\!&\!\! \cellcolor[rgb]{0.95,0.95,0.95} 0.00\% \!\!&\!\! \cellcolor[rgb]{0.95,0.95,0.95} 58.12\% \!\!&\!\! \cellcolor[rgb]{0.95,0.95,0.95} 0.00\% \!\!&\!\! \cellcolor[rgb]{0.95,0.95,0.95} 0.12\% \!\!&\!\! \cellcolor[rgb]{0.95,0.95,0.95} 15.23\% \!\!&\!\! \cellcolor[rgb]{0.95,0.95,0.95} 0.00\% \!\!&\!\! \cellcolor[rgb]{0.95,0.95,0.95} 0.00\% \!\!&\!\! \cellcolor[rgb]{0.95,0.95,0.95} 0.00\% \!\!&\!\! \cellcolor[rgb]{0.95,0.95,0.95} 5.53\% \!\!&\!\! \cellcolor[rgb]{0.95,0.95,0.95} 0.00\% \!\!&\!\! \cellcolor[rgb]{0.95,0.95,0.95} 27.83\% \!\!&\!\! \cellcolor[rgb]{0.95,0.95,0.95} 0.00\% \!\\
\cline{2-15}\\[-10pt]
& Error rate &\! 0.0114 \!\!&\!\! 0.0018 \!\!&\!\! 0.1648 \!\!&\!\! 0.0020 \!\!&\!\! 0.0248 \!\!&\!\! 0.0740 \!\!&\!\! 0.0069 \!\!&\!\! 0.0024 \!\!&\!\! 0.0988 \!\!&\!\! 0.1266 \!\!&\!\! 0.0247 \!\!&\!\! 0.0184 \!\!&\!\! 0.0134 \!\\
\cline{1-15}\\[-10.3pt]
\multirow{1.1}{*}{MTL w/ soft labels}& \cellcolor[rgb]{0.95,0.95,0.95} F-score &\! \cellcolor[rgb]{0.95,0.95,0.95} \textbf{11.08\%} \!\!&\!\! \cellcolor[rgb]{0.95,0.95,0.95} 0.00\% \!\!&\!\! \cellcolor[rgb]{0.95,0.95,0.95} \textbf{74.60\%} \!\!&\!\! \cellcolor[rgb]{0.95,0.95,0.95} \textbf{2.91\%} \!\!&\!\! \cellcolor[rgb]{0.95,0.95,0.95} \textbf{9.47\%} \!\!&\!\! \cellcolor[rgb]{0.95,0.95,0.95} \textbf{42.20\%} \!\!&\!\! \cellcolor[rgb]{0.95,0.95,0.95} \textbf{0.35\%} \!\!&\!\! \cellcolor[rgb]{0.95,0.95,0.95} 0.00\% \!\!&\!\! \cellcolor[rgb]{0.95,0.95,0.95} 0.80\% \!\!&\!\! \cellcolor[rgb]{0.95,0.95,0.95} \textbf{48.88\%} \!\!&\!\! \cellcolor[rgb]{0.95,0.95,0.95} 0.62\% \!\!&\!\! \cellcolor[rgb]{0.95,0.95,0.95} 46.60\% \!\!&\!\! \cellcolor[rgb]{0.95,0.95,0.95} 0.00\% \!\\\cline{2-15}
\multirow{0.9}{*}{($\beta = 1.0, T = 1.0$)}&&&&&&&&&&&&&\!\\[-10pt]
& Error rate &\! \textbf{0.0112} \!\!&\!\! 0.0018 \!\!&\!\! \textbf{0.1363} \!\!&\!\! \textbf{0.0019} \!\!&\!\! \textbf{0.0243} \!\!&\!\! 0.0610 \!\!&\!\! \textbf{0.0068} \!\!&\!\! 0.0024 \!\!&\!\! \textbf{0.0986} \!\!&\!\! 0.1034 \!\!&\!\! 0.0247 \!\!&\!\! 0.0157 \!\!&\!\! 0.0134\!\\[-1pt]
\wcline{1-15}
\end{tabular}
}
\vspace{-5pt}
\end{table*}
\vspace{-5pt}
\section{Evaluation Experiments}
\label{sec:experiments}
\vspace{-6pt}
\subsection{Experimental Conditions}
\label{ssec:conditions}
\vspace{-5pt}
To evaluate the performance of the proposed method for SED, we conducted event detection experiments.
For the evaluation, we constructed a dataset composed of parts of the TUT Sound Events 2016 development, 2017 development, and TUT Acoustic Scenes 2016 development \cite{Mesaros_EUSIPCO2016_01,Mesaros_DCASE2017_01}.
From the three datasets, we selected sound clips including four acoustic scenes, ``home,'' ``residential area,'' ``city center,'' and ``office,'' with a total duration of 192 minutes of audio.
The experimental data include the 25 types of sound event listed in Table~\ref{tbl:event_occurrence}.
Because the original TUT Acoustic Scenes 2016 development dataset does not have sound event annotations in the office scene, we annotated them using the same procedure as in \cite{Mesaros_EUSIPCO2016_01,Mesaros_DCASE2017_01}.
The sound event labels of the office scene annotated for this experiment are available in \cite{Imoto_dataset2019_01}.

As the input of networks, we used the 64-dimensional log mel-band energy, which has a frame length of 40 ms with 50\% overlap.
The acoustic features were input into the networks shown in Fig.~\ref{fig:propnet}, and active sound events were estimated by thresholding using an adaptive thresholding method \cite{Xu_DCASE2017_01}.
SED performance was evaluated using the segment-based F1-score and error rate \cite{Mesaros_AS2016_01}.
Other experimental conditions are listed in Table~\ref{tbl:parameter}.
\vspace{-11pt}
\subsection{Experimental Results}
\label{ssec:results}
\vspace{-4pt}
To obtain the results of each experiment, we trained each model and evaluated the detection performance with a four-fold cross-validation setup $\times$ 3 initial values of model parameters.
Table~\ref{tbl:performance01} and Fig.~\ref{fig:performance} show the average performance of sound event detection using CNN-BiGRU \cite{Cakir_TASLP2017_01} (hereafter referred to as CRNN), the MTL-based method \cite{Tonami_WASPAA2019_01} (hereafter referred to as MTL), and the proposed MTL-based method with the soft scene labels (hereafter referred to as MTL w/ soft labels) in terms of F-score and error rate.
The results show that when using the conventional MTL-based method, the F-score tend to decrease as the parameter $\alpha$ increases.
This means that because the conventional MTL-based method may not model the relationship between acoustic scenes and events precisely, a large amount of scene information leads to performance degradation.
On the other hand, the proposed method ($\beta=1.0, T=1.0$) improves the F-score of event detection performance by 3.80 percentage points compared with that of the conventional MTL-based method ($\alpha = 0.0001$).
Thus, the proposed method can utilize information on acoustic scenes for SED more effectively.

To examine the detection results in more detail, we list the event detection results for each event in Table~\ref{tbl:each_event}.
The results show that in many of the sound events, the proposed method achieves a higher F-score and error rate than the conventional methods.
For example, the proposed method can detect the acoustic events ``dishes'', ``fan,'' and ``people walking'' more accurately; the F-scores of these sound events increase by 10.28, 14.05, and 0.10 percentage points, respectively, compared with the conventional MTL-based method.
On the other hand, the event detection performance for ``(object) snapping,'' ``breathing,'' and ``wind blowing'' are not improved.
This may be because these sound events hardly occur in the recorded sound clips; thus, there is still a class imbalance problem between sound events, which will be addressed in the future.
%
%
%
%
\vspace{-13pt}
\section{Conclusion}
\vspace{-8pt}
In this paper, we proposed the SED method based on MTL of SED and ASC with soft scene labels.
In the proposed method, the teacher network for scene classification is first trained using hard scene labels, and the output of the trained teacher network is used as the soft scene label for training the student MTL-based network.
The experimental results obtained using the TUT Sound Events 2016, 2017, and TUT Acoustic Scenes 2016 datasets show that the proposed method outperforms the conventional MTL-based SED method by 3.80 percentage points in terms of the segment-based F1-score.
%
%
%
%
%
\vspace{-8pt}
\section{Acknowledgement}
\label{sec:ack}
\vspace{-7pt}
This work was supported by JSPS KAKENHI Grant Number JP19K20304 and NVIDIA GPU Grant Program.
%
%
%
\small
\bibliographystyle{IEEEbib}
\bibliography{IEEEabrv,ICASSP2020refs,KeisukeImoto09}
\end{document}